\documentstyle[epsf,aps,preprint]{revtex}
\begin{document}
\draft
\tightenlines
\title{Dressed-State Approach to Population Trapping in the Jaynes-Cummings
Model}
\author{D. Jonathan\thanks{%
Present address: Optics Section, The Blackett Laboratory, Imperial College,
London SW7 2BZ, United Kingdom; e-mail: d.jonathan@ic.ac.uk}, K. Furuya\thanks{e-mail:
furuya@ifi.unicamp.br}, A. Vidiella-Barranco\thanks{e-mail: vidiella@ifi.unicamp.br} \\
Instituto de F\'{\i}sica ``Gleb Wataghin'', Universidade Estadual de
Campinas, CEP 13083-970, Campinas, Brazil.}
\maketitle
\date{today}
\begin{abstract}
The phenomenon of atomic population trapping in the Jaynes-Cummings Model is
analysed from a dressed-state point of view. A general condition for the
occurrence of partial or total trapping from an arbitrary, pure initial
atom-field state is obtained in the form of a bound to the variation of the
atomic inversion. More generally, it is found that in the presence of
initial atomic or atom-field coherence the population dynamics is governed
not by the field's initial photon distribution, but by a `weighted
dressedness' distribution characterising the joint atom-field state. In
particular, individual revivals in the inversion can be analytically
described to good approximation in terms of that distribution, even in the
limit of large population trapping. This result is obtained through a
generalisation of the Poisson Summation Formula method for analytical
description of revivals developed by Fleischhauer and Schleich [Phys. Rev. A 
{\bf 47}, 4258 (1993)].
\end{abstract}

\section{Introduction}

One of the most fundamental models of quantum optical resonance is the
interaction of a single two-level atom with a single quantised mode of
radiation, described by the Jaynes-Cummings Hamiltonian \cite
{Jaynes63,Shore93}. Despite being simple enough to be analytically soluble
in the rotating-wave approximation, this model has been a long-lasting
source of insight into the nuances of the interaction between light and
matter. It has led to nontrivial predictions, such as the existence of
`collapses' and `revivals' in the atomic excitation \cite{Eberly80}, and
has also allowed a deeper understanding of the dynamical entangling and
disentangling of the atom-field system in the course of time \cite
{Phoenix88}. Further interest in the Jaynes-Cummings model (JCM) comes from
the fact that these predictions are directly accessible to experimental
verification. A JCM interaction can be experimentally realized in cavity-QED
setups \cite{Haroche94} and also, as an effective interaction, in
laser-cooled trapped ions \cite{Blockley92} (in which case the ionic
harmonic motion assumes the role normally played by the field mode). For
example, revivals in the atomic excitation have recently been observed in a
cavity-QED experiment \cite{Brune96}, providing direct evidence for the
discreteness of photons.

In spite of these successes, a closed analytical description of the
collapse-revival pattern has so far proved to be elusive; however, an
elegant approximation scheme valid for a number of initial conditions has
been presented by Fleischhauer and Schleich \cite{Fleischhauer93}, improving
the earlier work of Eberly and co-workers \cite{Eberly80}. Among other
things, they have demonstrated that, when the atom is initially completely
excited or de-excited, and the initial photon-number distribution of the
field ($P_{n}$) is sufficiently smooth (as is the case in a coherent state)
then the shape of each revival is a direct reflection of the shape of $P_{n}$%
. This direct relationship can be affected by the presence of initial atomic
coherence. It has been noticed \cite{Zaheer89} that if the atom is initially
prepared in a coherent superposition of its energy eigenstates, then the
revivals can be largely suppressed, effectively freezing the value of the
atomic state populations. Appropriately, these initial atomic states have
been referred to as ``atomic trapping states'' \cite
{Zaheer89,GeaBanacloche91}{\em . }This population trapping has been
connected to the existence of a specific phase difference between the
relative phase of the atomic superposition and the phase of the initial
field state \cite{Zaheer89}. In particular, in the resonant case exact
trapping has been shown to exist \cite{Cirac90} if the field is initially
prepared in a phase coherent state \cite{Shapiro91} (eigenstate of the
Susskind-Glogower phase operator \cite{Susskind64}). As has been pointed out
by Cirac and S\'{a}nchez-Soto \cite{Cirac91}, this can be understood by the
fact that in these cases only one state out of each pair of dressed
eigenstates of the JCM is ever populated.

We now note that, since the atomic excitation dynamics can be modified
merely by altering the initial atomic state, then it is apparent that the
initial photon distribution cannot be the sole responsible for the existence
and shape of revivals. In fact, as we demonstrate in this paper, the key to
understand the collapse-revival pattern under more general initial
conditions is to consider not only the initial coherence of the atom and
field by themselves, but also their joint properties as a single quantum
system. This is so even if the system is initially disentangled. Building on
Cirac and S\'{a}nchez-Soto's observation, we find that the revival structure
depends{\em \ }essentially{\em \ }on the {\it relative weight} of each
dressed eigenstate in the initial atom-field state. We{\em \ }are able to
identify an atom-field variable which plays, in the general case, the same
role as that of the photon distribution when the atom is initially excited
or de-excited. This allows us to estimate for any initial condition the
degree of suppression of the revivals (that is, the amount of trapping which
occurs). In addition, in some particular cases we are able to describe
analytically to a good approximation the shape of the revivals, even when
they are partially suppressed. To illustrate the dressed-state distribution
dependence of the revival patterns, we show that essentially identical
examples can be found both in the presence of initial atomic coherence, and
in cases where this coherence is not only atomic, but where atom and field
are already entangled.

Our work is organised as follows: in section II, we rewrite the dynamics of
the Jaynes-Cummings model from a point of view based on the entangled
dressed-state eigenbasis. With the help of an appropriate coordinate system,
the expression for the atomic inversion assumes a simpler form; as a
consequence, we are able to calculate bounds for the evolution of this
quantity for arbitrary initial conditions. In order to illustrate our
technique, specific cases are treated in section III. In section IV, we
generalise Fleischhauer and Schleich's scheme to include the cases where
trapping occurs. Our conclusions are summarised in section V.

\section{Dressed-State Coordinates for the JCM}

\subsection{The Model\label{modelsec}}

The Jaynes-Cummings Hamiltonian, on resonance and in the rotating-wave
approximation is given by \cite{Jaynes63,Shore93}: 
\begin{equation}
H=\frac{1}{2}\hbar \omega \hat{\sigma}_{z}+\hbar \omega \hat{a}^{\dagger
}\hat{a}+\hbar \lambda (\hat{\sigma}_{+}\hat{a}+\hat{a}^{\dagger }\hat{%
\sigma}_{-})
\end{equation}
where $\omega $ is the atomic transition frequency, $\lambda $ is the
atom-field dipole coupling constant, $\hat{a}^{\dagger }(\hat{a})$ is the
field photon creation (annihilation) operator and $\hat{\sigma}_{z},\hat{%
\sigma}_{\pm }$ are the atomic inversion, rising and lowering operators. The
first two terms in this expression describe the internal energy levels of
the spin-1/2-like 2-level atom and the harmonic oscillator-like field mode,
and the interaction term can be straightforwardly understood as a
one-excitation exchange between these two systems.{\it \ }

An important feature of this fully-quantised matter-radiation interaction is
that its steady states (known as `dressed states') are{\em \ }{\it %
entangled. }Switching to an interaction-picture representation for
convenience, it can be shown that these states and their corresponding
energy levels are \cite{Meystre91}: 
\begin{mathletters}
\begin{eqnarray}
\left| n\pm \right\rangle &=&\frac{1}{\sqrt{2}}\left[ \left|
e,n\right\rangle \pm \left| g,n+1\right\rangle \right]  \label{dressedstates}
\\
E_{n\pm } &=&\pm \hbar \lambda \sqrt{n+1}\equiv \pm \frac{\hbar }{2}%
\Omega _{n}  \label{energies}
\end{eqnarray}
($\Omega _{n}$ is the Rabi frequency). It is apparent that in all these
cases the atomic state is completely undetermined, being an equal mixture of
the excited and ground states. Thus, if we assume that at $t=0$ the atom and
field are independently prepared in the states 
\end{mathletters}
\begin{eqnarray}
\left| \psi \left( 0\right) \right\rangle _{A} &=&p\left| e\right\rangle
+q\left| g\right\rangle \\
\left| \Phi \left( 0\right) \right\rangle _{F} &=&\sum_{n=0}^{\infty
}c_{n}\left| n\right\rangle
\end{eqnarray}
then the initial state $\left| \Psi \left( 0\right) \right\rangle
_{AF}=\left| \psi \left( 0\right) \right\rangle _{A}\otimes \left| \Phi
\left( 0\right) \right\rangle _{F}$ will evolve into 
\begin{eqnarray}
\left| \Psi \left( t\right) \right\rangle _{AF} &=&qc_{0}\left|
g,0\right\rangle +  \nonumber \\
&&+\sum_{n=0}^{\infty }\left[ 
\begin{array}{c}
pc_{n}\cos \left( \lambda t\sqrt{n+1}\right) - \\ 
-iqc_{n+1}\sin \left( \lambda t\sqrt{n+1}\right)
\end{array}
\right] \left| e,n\right\rangle +  \nonumber \\
&&+\sum_{n=0}^{\infty }\left[ 
\begin{array}{c}
qc_{n+1}\cos \left( \lambda t\sqrt{n+1}\right) - \\ 
-ipc_{n}\sin \left( \lambda t\sqrt{n+1}\right)
\end{array}
\right] \left| g,n+1\right\rangle  \label{interevol}
\end{eqnarray}

Despite being straightforwardly solvable in this way, the JCM is well-known
for the fact that the time-evolution of most expectation values is usually
expressible only in series form. For instance, assume that at time $t=0$ we
have: 
\begin{eqnarray}
\left| \Psi \left( 0\right) \right\rangle &=&\left( p\left| e\right\rangle
+q\left| g\right\rangle \right) \otimes \left( \sum_{n=0}^{\infty
}c_{n}\left| n\right\rangle \right)  \nonumber \\
&=&qc_{0}\left| g,0\right\rangle +\sum_{n=0}^{\infty }pc_{n}\left|
e,n\right\rangle +qc_{n+1}\left| g,n+1\right\rangle  \label{t=0fact}
\end{eqnarray}
The time evolution of the atomic population inversion is then given by:

\begin{eqnarray}
\left\langle \hat{\sigma}_{z}\right\rangle \left( t\right)
&=&2\sum_{n=0}^{\infty }\left| pc_{n}\cos \left( \lambda t\sqrt{n+1}\right)
-iqc_{n+1}\sin \left( \lambda t\sqrt{n+1}\right) \right| ^{2}-1  \nonumber \\
&=&-\left| qc_{0}\right| ^{2}+\sum_{n=0}^{\infty }\left( \left|
pc_{n}\right| ^{2}-\left| qc_{n+1}\right| ^{2}\right) \cos \left( 2\lambda t%
\sqrt{n+1}\right) +  \nonumber \\
&&+2{\it Im} \left( \left( pc_{n}\right) ^{*}qc_{n+1}\right) \sin \left(
2\lambda t\sqrt{n+1}\right)  \label{complinver}
\end{eqnarray}

where 
\begin{equation}
\sum_{n=0}^{\infty }\left| pc_{n}\right| ^{2}+\left| qc_{n+1}\right|
^{2}=1-\left| qc_{0}\right| ^{2}
\end{equation}
In particular, if at $t=0$ the atom is completely excited $(q=0)$ or in the
ground state $(p=0)$, and admitting for simplicity that $c_{0}$ is
negligible, then this expression reduces to the simpler form 
\begin{equation}
\left\langle \sigma _{z}\right\rangle \left( t\right) =\pm
\sum_{n=0}^{\infty }P_{n}\cos \left( 2\lambda \sqrt{n+1}t\right)
\label{invfactcase}
\end{equation}
where $P_{n}=\left| c_{n}\right| ^{2}$is the initial photon distribution of
the field. Due to the $\sqrt{n}$ scaling of the various Rabi frequencies in
this sum, no closed form for it is known. Nevertheless, it is apparent that
features of this time evolution, such as collapse-revival phenomena, must be
directly related to the characteristics of $P_{n}$. The
Fleischhauer-Schleich approximation scheme, which we shall discuss in
section IV, allows this relationship to be explicitly described in certain
situations, such as when $P_{n}$ is sufficiently smooth.

Apart from the simple case represented above, in general there is no way to
relate the inversion dynamics exclusively to the initial field state. As can
be seen from expression (\ref{complinver}), it will usually depend on atomic
and field variables in an apparently complicated way. In the next section,
we show how this evolution can be recast in a simple form, reminiscent of
expression (\ref{invfactcase}), independently of the initial conditions.

\subsection{Dressed-State Coordinates}

Our discussion will be based on the relative contribution of each dressed
state for the global atom-field state. For simplicity, we restrict ourselves
to {\it pure }initial conditions. Therefore, we may expand an arbitrary
initial state in terms of $\left| n\pm \right\rangle :$

\begin{equation}
\left| \Psi (0)\right\rangle =w_{-1}\left| g;0\right\rangle
+\sum_{n=0}^{\infty }w_{n}e^{i\chi _{n}}\left| \psi _{n}\right\rangle ,
\label{inista}
\end{equation}
where 
\begin{equation}
\left| \psi _{n}\right\rangle =\left[ \cos \left( \frac{\theta _{n}}{2}%
\right) \left| n+\right\rangle +e^{-i\phi _{n}}\sin \left( \frac{\theta _{n}%
}{2}\right) \left| n-\right\rangle \right] ,  \label{dresscomp}
\end{equation}
with the normalisation condition $\sum_{n=-1}^{\infty }w_{n}^{2}=1.$ The
parameters $w_{n}\in \left[ 0,1\right] ,$ $\theta _{n}\in \left[ 0,\pi
\right] $ and $\chi _{n},\phi _{n}\in \left[ 0,2\pi \right] ,$ will
henceforth be referred to as {\it dressed-state coordinates}.

The use of this coordinate system allows us to obtain a simple geometrical
picture of the manifold of pure atom-field states. For each value of $%
n\geq 0$ one can imagine a Bloch-type spherical shell corresponding to
the two-level system formed by $\left| n\pm \right\rangle $ (Fig. 1), which 
are parametrised by the spherical coordinates $\theta _n$
and $\phi _n.$ The poles $\left( \theta _n=0,\pi \right) $ are associated to
the dressed states $\left| n\pm \right\rangle $, and points on the equator $%
\left( \theta _n=\frac \pi 2\right) $ to the states of the form $\frac 1{%
\sqrt{2}}\left[ \left| n+\right\rangle +e^{-i\phi _n}\left| n-\right\rangle
\right] $, which include the particular product states $\left|
e,n\right\rangle $ and $\left| g,n+1\right\rangle $ (corresponding to $\phi
_n=0,\pi $). Therefore the coordinate $\theta _n$ gives a measure of what we
shall call the `{\it dressedness}' (or degree of proximity to the nearest
dressed state) of the components $\left| \psi _n\right\rangle $ of $\left|
\Psi (0)\right\rangle $. We note that, while this property is related to the 
{\it entanglement }of $\left| \psi _n\right\rangle $ (since the dressed
states are in fact maximally entangled), they are two different concepts:
states with the same dressedness, such as $\left| e,n\right\rangle $ and $%
\cos \mu \left| e,n\right\rangle -i\sin \mu \left| g,n+1\right\rangle $ can
have different amounts of entanglement (as measured by the von Neumann entropy of
their subsystems). Meanwhile, the weight factors $w_n$ measure the relative
importance of each of these components. In physical terms, their squares $%
w_n^2$ correspond to the probability distribution for measurements of the 
{\it total excitation number }operator{\it \ } 
\begin{equation}
\hat{X}_{tot}=\frac 12\hat{\sigma}_z+\hat{N}.
\end{equation}
{\em \ }on the initial state $\left| \Psi \left( 0\right) \right\rangle $ ( $%
w_n^2$ corresponds to the probability for $n+1$ excitations). We note that,
in the special case where the atom is prepared in one of the states $\left|
e\right\rangle $ or $\left| g\right\rangle $, then $\theta _n\equiv \frac
\pi 2$ and $\phi _n\equiv 0\left( \pi \right) $ for all $n$, while $w_n^2$
reduces to the photon number distribution: 
\begin{equation}
w_n^2\rightarrow \left\{ \begin{array}{c}
P_n=\left| \left\langle n|\Phi \left(
0\right) \right\rangle _F\right| ^2\text{ (if }\left| e\right\rangle )\\
P_{n+1}=\left| \left\langle n+1|\Phi \left( 0\right) \right\rangle _F\right|
^2\text{ (if }\left| g\right\rangle )
\end{array}\right. .
\end{equation}

We now show that the dressed-state coordinate system allows a simple
description of the time evolution of any initial state. From equations (\ref
{dressedstates},b), the evolution of state (\ref{inista}) is given by

\begin{equation}
\left| \Psi \left( t\right) \right\rangle =w_{-1}\left| g,0\right\rangle
+\sum_{n=0}^\infty w_ne^{\frac i2\left( 2\chi _n-\Omega _nt\right) }\left[
\cos \left( \frac{\theta _n}2\right) \left| n+\right\rangle +e^{-i\left(
\phi _n-\Omega _nt\right) }\sin \left( \frac{\theta _n}2\right) \left|
n-\right\rangle \right] .
\end{equation}
We can re-express this directly in terms of the dressed-state coordinates:

\begin{eqnarray}
w_{n}(t) &=&w_{n}(0)  \nonumber \\
\theta _{n}\left( t\right) &=&\theta _{n}\left( 0\right) \\
\chi _{n}\left( t\right) &=&\chi _{n}\left( 0\right) -\frac{1}{2}\Omega _{n}t
\nonumber \\
\phi _{n}\left( t\right) &=&\phi _{n}\left( 0\right) -\Omega _{n}t  \nonumber
\end{eqnarray}
Thus, $w_{n}$ and $\theta _{n}$ are both constants of motion, while the
angles $\chi _{n}$ and $\phi _{n}$ precess with a constant angular velocity
proportional to the Rabi frequency $\Omega _{n}$. A way of visualising this
is by means of a geometrical image as in Fig. 1. The simplicity of this
picture reflects the symmetries of the resonant JCM: the conservation of $%
\hat{X}_{tot}$ and the fact that, in the interaction picture, the energies
in each 2-dimensional eigensubspace generated by $\left| n\pm \right\rangle $
are always symmetric with respect to zero.

\subsubsection{Population Inversion in Dressed-State Coordinates\label%
{inversec}}

Let us now apply this coordinate system to describe the evolution of atomic
variables.{\em \ }In the basis $\left\{ \left| e\right\rangle ,\left|
g\right\rangle \right\} ,$ the reduced atomic density operator can be
expressed as: 
\begin{equation}
\rho _{A}\left( t\right) =\left[ 
\begin{array}{ll}
\rho _{ee}\left( t\right) & \rho _{eg}\left( t\right) \\ 
\rho _{eg}^{*}\left( t\right) & \rho _{gg}\left( t\right)
\end{array}
\right]  \label{atomDM}
\end{equation}
where

\begin{mathletters}
\begin{eqnarray}
\rho _{ee}\left( t\right) &=&\frac{1}{2}\left( 1-w_{-1}^{2}\right) +\frac{1}{%
2}\sum_{n=0}^{\infty }w_{n}^{2}\sin \left( \theta _{n}\right) \cos \left(
\phi _{n}\left( t\right) \right) ;\;\;\;\;\;\;\rho _{gg}\left( t\right)
=1-\rho _{ee}\left( t\right)  \label{rhoee(t)} \\
\rho _{eg}\left( t\right) &=&\frac{w_{-1}w_{0}}{\sqrt{2}}e^{i\left( \chi
_{0}\left( t\right) \right) }\left( \cos \left( \frac{\theta _{0}}{2}\right)
+e^{-i\phi _{0}\left( t\right) }\sin \left( \frac{\theta _{0}}{2}\right)
\right) +  \nonumber \\
&&+\frac{1}{2}\sum_{n=0}^{\infty }w_{n}w_{n+1}e^{i\left( \chi _{n+1}\left(
t\right) -\chi _{n}\left( t\right) \right) }\left( \cos \left( \frac{\theta
_{n+1}}{2}\right) +e^{-i\phi _{n+1}\left( t\right) }\sin \left( \frac{\theta
_{n+1}}{2}\right) \right) \times  \nonumber \\
&&\times \left( \cos \left( \frac{\theta _{n}}{2}\right) -e^{i\phi
_{n}\left( t\right) }\sin \left( \frac{\theta _{n}}{2}\right) \right)
\label{rhoeg(t)}
\end{eqnarray}

We thus see that the evolution of the atomic level populations, and
therefore of the population inversion, can be expressed in a simple form for
any initial condition: 
\end{mathletters}
\begin{equation}
\left\langle \hat{\sigma}_z\left( t\right) \right\rangle =2\rho _{ee}\left(
t\right) -1=-w_{-1}^2+\sum_{n=0}^\infty w_n^2\sin \left( \theta _n\right)
\cos \left( \phi _n\left( t\right) \right) .  \label{popinv}
\end{equation}
This should be contrasted with the much more complicated expression (\ref
{complinver}) for this same quantity, written in terms of separate atomic
and field coordinates. In fact, we can see that the present expression is a
direct generalisation of eq. (\ref{invfactcase}), obtained by phase-shifting
the $n^{th}$ term in the sum by $\phi _n\left( 0\right) $ and by replacing
the photon number distribution $P_n$ with 
\begin{equation}
D_n=w_n^2\sin \left( \theta _n\right) .  \label{weidist}
\end{equation}
It is not difficult to understand the reason for this substitution. First of
all, as was noted by Cirac and S\'{a}nchez-Soto \cite{Cirac91}, dressed
states have no population inversion. Therefore, if a component $\left| \psi
_n\right\rangle $ of a given state $\left| \Psi \left( 0\right)
\right\rangle $ has a large `dressedness' ($\sin \left( \theta _n\right)
\rightarrow 0$), it will contribute very little to the overall inversion$.$
Moreover, even if $\left| \psi _n\right\rangle $ does have a large
inversion, this may still be of little consequence to the total average $%
\left\langle \hat{\sigma}_z\left( t\right) \right\rangle $ if its relative
importance $w_n^2$ is small. Thus, the product $D_n,$ which we shall refer
to as the `weighted dressedness distribution' gives the appropriate
magnitude of the contribution of $\left| \psi _n\right\rangle $ to the
inversion. In other words, for general initial conditions it is this
distribution, not $P_n$, that will control the evolution of the inversion.
For instance, we can expect the existence of collapse-revival structures if $%
D_n$ is `narrow', in the sense that few values of $n$ (and therefore few
Rabi frequencies $\Omega _n)$ feature significantly in the sum.

\subsubsection{Population Trapping\label{trap}}

The main difference between the general evolutions allowed by equation (\ref
{popinv}) and those obtained in the special case (\ref{invfactcase}) lies in
the fact that the distribution $P_{n}$ is normalised{\it ,} while $D_{n}$ is
generally {\it not}. This implies that, in general, there will be an upper
bound for the allowed range of variation of the population inversion, given
by:

\begin{equation}
\left| \left\langle \hat{\sigma}_{z}\left( t\right) \right\rangle
+w_{-1}^{2}\right| \leq \sum_{n=0}^{\infty }D_{n}\equiv M\leq 1
\label{bound}
\end{equation}
As we can see, this bound is specified essentially by the {\it average
dressedness }of the components of $\left| \Psi \left( 0\right) \right\rangle 
$. For example: if the most important components (those with largest weights 
$w_{n}^{2}$) are highly dressed ($\sin \left( \theta _{n}\right) $ is
small), then the amplitude of variation of the population inversion will
also be small ($M\sim 0$).{\em \ }In other words, these are the conditions
for the existence of {\it population trapping }in the (resonant) JCM{\it . }%
The steady-state value of the inversion is given simply by $-w_{-1}^{2}$,
i.e., by the fixed population in the ground state. We emphasise that, given
any initial condition, the bound $M$ can be immediately calculated from the
constants of the motion $w_{n},$ $\theta _{n}$, thus giving an estimate of
the amount of trapping that can be expected from that state's evolution \cite
{Hillery87}.

\section{Examples}

In order to illustrate our technique, we analyse three classes of initial
atom-field states for which population trapping occurs:

\subsection{Perfect Trapping States}

In \cite{Cirac90} , Cirac and S\'{a}nchez-Soto found a class of factorised
initial conditions which exhibit {\it perfect} trapping, that is, for which
the atomic populations are strictly constant over time. These were of the
form:

\begin{equation}
\left| \Psi \left( 0\right) \right\rangle =\sqrt{1-\left| z^{2}\right| }%
\left( z\left| e\right\rangle +\left| g\right\rangle \right) \otimes \frac{1%
}{\sqrt{1+\left| z^{2}\right| }}\sum_{n=0}^{\infty }z^{n}\left|
n\right\rangle   \label{exact}
\end{equation}
where $\left| z\right| <1.$ In this case, the field is prepared in an
eigenstate of the Susskind-Glogower phase operator $\hat{V}%
=\sum_{n=0}^{\infty }\left| n\right\rangle \left\langle n+1\right| $ \cite
{Susskind64} (`phase-coherent state'\cite{Shapiro91}), while the atomic
superposition is chosen so that the phase of the atomic dipole matches that
of the eigenvalue $z$ of the field state \cite{remark}. The resulting
population trapping can thus be attributed to a suitable matching of atomic
and field parameters. However, as was later noticed by the same authors \cite
{Cirac91}, a clearer explanation can be found by using a dressed-state point
of view: Rewriting this state in terms of the dressed-state basis, we find

\begin{equation}
\left| \Psi \left( 0\right) \right\rangle =\sqrt{\frac{1-\left| z^2\right| }{%
1+\left| z^2\right| }}\left[ \left| g,0\right\rangle +\sum_{n=0}^\infty 
\sqrt{2}z^n\left| n+\right\rangle \right] .  \label{exact2}
\end{equation}
The reason for the existence of trapping now becomes immediately clear: all
the components $\left| \psi _n\right\rangle $ of this state are completely
dressed ($\sin \left( \theta _n\right) \equiv 0$), and thus do not
contribute to the inversion.

Cirac and S\'{a}nchez-Soto also stated \cite{Cirac90} that these are the 
{\it only }factorised states exhibiting perfect trapping. We note that this
is not strictly true: the same result is still achieved for all states of
the form 
\begin{equation}
\left| \Psi \left( 0\right) \right\rangle =\sqrt{1-\left| z^{2}\right| }%
\left( z\left| e\right\rangle +\left| g\right\rangle \right) \otimes \frac{1%
}{\sqrt{1+\left| z^{2}\right| }}\sum_{n=0}^{\infty }j\left( n\right)
z^{n}\left| n\right\rangle  \label{Exact3}
\end{equation}
where $j\left( n\right) $ can be $\pm 1$ and $\left| z\right| <1$ (in this
case, the field state is not in general an eigenstate of $\hat{V}$). Once
again, this can be seen by rewriting the state in the dressed-state basis.
In this case, we obtain 
\begin{equation}
\left| \Psi \left( 0\right) \right\rangle =\sqrt{\frac{1-\left| z^{2}\right| 
}{1+\left| z^{2}\right| }}\left[ j\left( 0\right) \left| g,0\right\rangle
+\sum_{n=0}^{\infty }\sqrt{2}j\left( n\right) z^{n}\left| n\pm \right\rangle
\right]
\end{equation}
where for each $n$ there is a single dressed state $\left| n\pm
\right\rangle $ present, whose sign is the same as that of the ratio $\frac{%
j\left( n+1\right) }{j\left( n\right) }.$ The converse statement is also
easy to show: suppose $\left| \Psi \left( 0\right) \right\rangle $ is a
perfect trapping state, of the form:

\begin{eqnarray}
\left| \Psi \left( 0\right) \right\rangle &=&k\left( z\left| e\right\rangle
+\left| g\right\rangle \right) \otimes \sum_{n=0}^{\infty }a_{n}\left|
n\right\rangle  \nonumber \\
&=&ka_{0}\left| g,0\right\rangle +\sum_{n=0}^{\infty }za_{n}\left|
e,n\right\rangle +a_{n+1}\left| g,n+1\right\rangle
\end{eqnarray}
where $k$ is a constant and $a_{n},$ $z$ are arbitrary. Then for each $%
\left| \psi _{n}\right\rangle $ component of this state to be perfectly
dressed it is necessary that: 
\begin{equation}
a_{n+1}=\pm za_{n}
\end{equation}
and hence 
\begin{equation}
a_{n}\propto \pm z^{n}.
\end{equation}
We recover expression (\ref{Exact3}) by noting that $\left| z\right| $ must
be less than $1$ for $\left| \Psi \left( 0\right) \right\rangle $ to be
normalisable. Thus, for the resonant JCM\ there can be no population
trapping with positive population inversion.

\subsection{Zaheer-Zubairy `atomic trapping states'}

While the initial conditions considered above lead to perfect trapping,
preparing the appropriate field states requires elaborate state-engineering
techniques. It has been shown (numerically) by Zaheer and Zubairy, however,
that {\it approximate} population trapping is also possible if the cavity
field is initially in a coherent state $\left| \alpha \right\rangle $, which
are easily accessible in experiment \cite{Zaheer89}. This is portrayed in
Fig. 2: if the atom is initially completely excited, the
inversion evolves according to the familiar collapse-revival pattern \cite
{Eberly80}; however, by suitably rotating the atomic state, the revival
peaks become more and more reduced, practically disappearing when the atom
is prepared in state 
\begin{equation}
\left| \psi \left( 0\right) \right\rangle _A=\frac 1{\sqrt{2}}\left( \left|
e\right\rangle +e^{-i\nu _a}\left| g\right\rangle \right)  \label{zz}
\end{equation}
(where $\nu _\alpha $ is the phase of the field's coherent amplitude $\alpha 
$). Close inspection also reveals that, in this limit, the revival peaks
become split into doublets.

In their work, Zaheer and Zubairy gave no quantitative explanation for this
quenching of the revivals, attributing it qualitatively to a ``destructive
interference between the atomic dipole and the cavity eigenmode ''. Later,
Cirac and S\'{a}nchez-Soto pointed out that the existence of approximate
trapping in this case{\em \ }should be expected due to similarities between
the properties of coherent states and of the phase coherent field states 
\cite{Cirac90}.

With the help of our dressed-state coordinate system, we can now give a
quantitative description of this effect. Let us suppose that the atom-field
system is prepared in the state:

\begin{equation}
\left| \Psi _{ZZ}\left( \alpha ,\gamma ,\xi \right) \right\rangle =\left[
\cos \gamma \left| e\right\rangle +e^{-i\xi }\sin \gamma \left|
g\right\rangle \right] \otimes \left| \alpha \right\rangle ,  \label{zzstate}
\end{equation}
where the atom is in a coherent superposition of ground and excited states
and the field is in a coherent state $\left( \alpha =|\alpha |e^{i\nu
_{\alpha }}\right) $. Expanding this state in terms of the dressed-state
basis, we find that its dressed-state coordinates have the following values
(we have only listed those relevant for the atomic inversion):

\begin{eqnarray}
w_{n}^{2}\left( \alpha ,\gamma ,\xi \right) &=&\frac{|\alpha
|^{2n}e^{-\left| \alpha \right| ^{2}}}{\left( n+1\right) !}\left[ \left(
n+1\right) \cos ^{2}(\gamma )+|\alpha |^{2}\sin ^{2}(\gamma )\right] \text{ }%
,n\geq -1  \nonumber \\
\sin \theta _{n}\left( \alpha ,\gamma ,\xi \right) &=&\frac{\left| \left(
n+1\right) \cos ^{2}(\gamma )-|\alpha |^{2}e^{i2\left( \nu _{\alpha }-\xi
\right) }\sin ^{2}(\gamma )\right| }{\left( n+1\right) \cos ^{2}(\gamma
)+|\alpha |^{2}\sin ^{2}(\gamma )}\;,\;\left( \theta _{n}\in \left[ 0,\pi
\right] \right)  \label{zzdc} \\
\cos \phi _{n}\left( \alpha ,\gamma ,\xi \right) &=&\frac{\left( n+1\right)
\cos ^{2}(\gamma )-|\alpha |^{2}\sin ^{2}(\gamma )}{\left| \left( n+1\right)
\cos ^{2}(\gamma )-|\alpha |^{2}e^{i2\left( \nu _{\alpha }-\xi \right) }\sin
^{2}(\gamma )\right| }  \nonumber \\
\sin \phi _{n}\left( \alpha ,\gamma ,\xi \right) &=&\frac{|\alpha |\sqrt{%
\left( n+1\right) }\sin (2\gamma )\sin \left( \nu _{\alpha }-\xi \right) }{%
\left| \left( n+1\right) \cos ^{2}(\gamma )-|\alpha |^{2}e^{i2\left( \nu
_{\alpha }-\xi \right) }\sin ^{2}(\gamma )\right| },\left( \phi _{n}\in
\left[ 0,2\pi \right] \right) .  \nonumber
\end{eqnarray}
The weight factors $w_{n}^{2}$ follow a Poisson distribution when $\gamma $
is equal to zero (they reduce to the photon number distribution of the
coherent field) and have small deviations from that distribution for any
value of $\gamma $. For moderately large $\left| \alpha \right| $,
therefore, $w_{n}^{2}$ is maximised around the integer $n_{\max }$ closest
to $\left| \alpha \right| ^{2}$ or $\left| \alpha \right| ^{2}-1.$
Meanwhile, it is not difficult to show that the $\sin \theta _{n}$
distribution has an absolute minimum, located around $n_{\min }\approx $ $%
\left| \alpha \right| ^{2}\tan ^{2}\gamma -1,$ with minimum value given by 
\begin{equation}
\sin \left( \theta _{n_{\min }}\right) \simeq \sin \left( \left| \nu
_{\alpha }-\xi \right| \right)
\end{equation}
This indicates the component of $\left| \Psi \left( \alpha ,\gamma ,\xi
\right) \right\rangle $ having the largest dressedness.

Following the procedure outlined in section \ref{trap}, we can now see that
to attain the highest possible degree of population trapping we must
maximise the largest dressedness, and match it with the largest weight
factor: 
\begin{mathletters}
\begin{eqnarray}
n_{\max } &=&n_{\min }\rightarrow \tan \gamma \simeq 1  \label{cond1} \\
\sin \left( \theta _{n_{\min }}\right) &=&0\rightarrow \nu _{\alpha }=\xi
\label{cond2}
\end{eqnarray}
We thus explain the ``atomic trapping state'' (\ref{zz}) found by Zaheer and
Zubairy: it is the one for which the total atom-field state $\left| \Psi
_{ZZ}\left( \alpha ,\gamma ,\xi \right) \right\rangle $ most closely
resembles a ``perfect trapping state'', in the sense that its most important
components $\left| \psi _{n}\right\rangle $ closely resemble dressed states.

It is clear from expression (\ref{zzdc}) that the degree of trapping
experienced from initial state $\left| \Psi _{ZZ}\left( \alpha ,\gamma ,\xi
\right) \right\rangle $ depends both on the relative weight $\gamma $ of
each atomic state and on the phase difference $\left( \nu _\alpha -\xi
\right) $ between the atomic dipole phase and the field's coherent
amplitude. In Fig. 3 we illustrate the change in the
weighted dressedness distribution $D_n$ as a function of these parameters:\
on the left-hand side, we take equal weights for $\left| e\right\rangle $
and $\left| g\right\rangle $ , satisfying condition (\ref{cond1}), and vary $%
\left( \nu _\alpha -\xi \right) $ from $\frac \pi 2$ to $0$. We can see the
quenching of the distribution as the dressedness of the component $\left|
\psi _{n_{\max }}\right\rangle $ is increased. On the right-hand side, we
keep the phase difference null and vary the relative weight from $0$ to $%
\frac \pi 4$, obtaining a similar effect (in both cases, we have chosen $%
\alpha =7$). Comparing this figure with Fig. 2, it is
apparent that there is a striking similarity between the profile of the $D_n$
distribution and the envelope of the revivals in the corresponding time
evolution, including the appearance of a doublet structure in the limit of
the optimal trapping conditions. As we shall see in section IV,
this is no coincidence.

\subsection{`Even-odd' entangled states}

The achievement of population trapping in the previous examples was seen to
be ultimately caused by the quenching of the $D_{n}$ distribution, and
therefore by the joint properties of the atom-field system. However, since
the initial states considered were factorised into atomic and field states,
it could also be argued that the trapping was due to a suitable matching of
independent atomic and field parameters (i.e., the phase of the field's
coherent amplitude matching the atomic dipole phase in the previous
example). In order to stress the underlying importance of the dressed-state
point of view, we now show that essentially identical evolution and trapping
patterns for the population inversion can be obtained even when the atom and
field cannot be considered as independent quantum systems.

Consider the following class of states of the atom-field system: 
\end{mathletters}
\begin{equation}
\left| \Psi _{EO}\left( \alpha ,\gamma ,\xi \right) \right\rangle =\cos
(\gamma )\left| e\right\rangle _A\left| even\right\rangle _f+\sin (\gamma
)e^{i\xi }\left| g\right\rangle _A\left| odd\right\rangle _f,
\label{evenodd}
\end{equation}
where $\left| even\right\rangle _f$\ $=\frac 1{\sqrt{2}}\left( \left| \alpha
\right\rangle +\left| -\alpha \right\rangle \right) $ and $\left|
odd\right\rangle _f=\frac 1{\sqrt{2}}\left( \left| \alpha \right\rangle
-\left| -\alpha \right\rangle \right) $ are respectively even and odd
coherent states of the cavity field (here $\left| \alpha \right| $ is
assumed large enough so that these states can be considered normalised). It
can be seen that, apart from the limiting cases where $\gamma \rightarrow
0,\frac \pi 2$, these states are {\it entangled}. Furthermore, since $\left|
even\right\rangle _f$\ $\left( \left| odd\right\rangle _f\right) $\ have
nonzero amplitudes only for even (odd) photon numbers, an expansion of this
state in terms of dressed states $\left| n\pm \right\rangle $ features only
those with {\it even }$n$ (in other words, $w_{2n-1}=0$). Physically, this
implies that states of this form have {\it zero} average electric and
magnetic fields, and {\it zero} average atomic polarisation. 
\begin{equation}
\left\langle \hat{a}+\hat{a}^{\dagger }\right\rangle =\left\langle \hat{a}-%
\hat{a}^{\dagger }\right\rangle =\left\langle \hat{\sigma}_x\right\rangle
=\left\langle \hat{\sigma}_y\right\rangle =0
\end{equation}
Thus, if at t = 0 the atom-filed system is in state $\left| \Psi _{EO}\left(
\alpha ,\gamma ,\xi \right) \right\rangle $, the population inversion
evolves according to eq. (\ref{popinv}), but with only the even terms
present: 
\begin{equation}
\left\langle \hat{\sigma}_z\left( t\right) \right\rangle =\sum_{n=0}^\infty
w_{2n}^2\sin (\theta _{2n})\cos (\phi _{2n}(t)).
\end{equation}

Now, it turns out that all these even-indexed dressed-state coordinates have
values virtually identical to the ones listed in expression (\ref{zzdc})
above for the state $\left| \Psi _{ZZ}\left( \alpha ,\gamma ,\xi \right)
\right\rangle $ with the same parameters $\alpha ,\gamma ,\xi .$ The only
difference is a multiplication of the weights $w_{2n}$ in the present case
by a factor of 2, to preserve normalisation in spite of the absence of
odd-indexed components. The resulting time-evolution of the atomic
inversion, plotted in Fig. 4, shows that the quenching of
revivals and appearance of a doublet structure as $\left( \nu _\alpha -\xi
\right) \rightarrow 0$ proceed in essentially the same manner as in the
previous case, the main difference being the doubling of the frequency of
revivals \cite{remark2}.{\em \ }We note, however, that in the present case
neither of the parameters $\nu _\alpha ,\xi ${\em \ }can be unambiguously
assigned exclusively to the atom or field, so that population trapping in
this case must necessarily be understood as a result of the joint atom-field
properties of the state in question.

It is also worth remarking that, due to the lack of atomic polarisation, the
atomic reduced density operators of these states are always diagonal in the $%
\left| e\right\rangle ,\left| g\right\rangle $ basis . Therefore, the {\it %
reduced entropy } $S_a=-Tr\rho _a\ln \rho _a$, which measures the degree of
entanglement between the atom and the field, is entirely determined by the
population inversion: 
\begin{eqnarray}
S_a\left( t\right) &=&-\rho _{ee}\left( t\right) \ln \left( \rho _{ee}\left(
t\right) \right) -\left( 1-\rho _{ee}\left( t\right) \right) \ln \left(
1-\rho _{ee}\left( t\right) \right)  \nonumber \\
&=&-\frac 12\left( 1-\left\langle \hat{\sigma}_z\left( t\right)
\right\rangle \right) \ln \left( \frac 12\left( 1-\left\langle \hat{\sigma}%
_z\left( t\right) \right\rangle \right) \right)  \nonumber \\
&&-\frac 12\left( 1+\left\langle \hat{\sigma}_z\left( t\right) \right\rangle
\right) \ln \left( \frac 12\left( 1+\left\langle \hat{\sigma}_z\left(
t\right) \right\rangle \right) \right) .
\end{eqnarray}
Thus, in this case the existence of population trapping is equivalent to the
atom and field remaining (nearly) maximally entangled during their entire
time-evolution. For any given state $\left| \Psi _{EO}\left( \alpha ,\gamma
,\xi \right) \right\rangle ,$ a lower bound to the value of $S_a\left(
t\right) $ at any given instant of the evolution is then given by: 
\begin{equation}
S_{\min }=-\frac 12\left( 1-M\right) \ln \left( \frac 12\left( 1-M\right)
\right) -\frac 12\left( 1+M\right) \ln \left( \frac 12\left( 1+M\right)
\right)  \label{sbound}
\end{equation}
where $M$ is the bound defined in eq. (\ref{bound}). For instance, in the
case of the `trapping state' depicted in Fig. 4, where, $\alpha =7$%
, $\gamma =\frac \pi 4,$ $\xi =0$ , we have $S_{\min }=0.69005$, very close
to the maximum possible value $\ln 2\simeq 0.69315.$ Finally, we note that
despite the existence of this lower bound, it is still possible to devise
schemes by which such entangled atom-field states can be constructed \cite
{Jonathan97,Jonathan98}.

\section{Poisson Summation Formula for revivals in the case of population
trapping}

In \cite{Fleischhauer93}, Fleischhauer and Schleich obtained approximate
analytical expressions for the evolution of the atomic inversion, using a
stationary-phase method based on the {\it Poisson Summation Formula} \cite
{Courant53}. They showed that in many cases the inversion can be written as
a sum in which each term $\omega _k\left( t\right) $ is non-negligible only
during a certain extension of time. This is in contrast for instance with
the expressions used in section \ref{modelsec} above, where each term in the
summation is periodic.

Their work was restricted to initial states of the form $\left|
g\right\rangle _{A}\otimes \left| \Phi \right\rangle _{F}$, in which the
inversion is given by eq. (\ref{invfactcase}). They were able to show that,
if the photon distribution $P_{n}$ of $\left| \Phi \right\rangle _{F}$ is
sufficiently smooth, then the $k^{th}$ term in the Poisson sum for the
inversion is given by

\begin{equation}
\omega _{k}\left( t\right) =-\frac{\lambda t}{\pi \sqrt{2k^{3}}}P\left( n=%
\frac{\lambda ^{2}t^{2}}{4\pi ^{2}k^{2}}\right) \cos \left( \frac{\lambda
^{2}t^{2}}{2\pi k}-\frac{\pi }{4}\right)
\end{equation}
where $P\left( n\right) $ is a continuous interpolation of $P_{n}$.

The main interest of this formulation is the fact that, if $P_n$ is also
sufficiently narrow, then the term $\omega _k\left( t\right) $ describes to
a high accuracy the evolution of the inversion during the $k^{th}$ revival.
This can be readily seen from the formula above: this term describes a rapid
oscillation in time, modulated by an envelope centred around $t\simeq \frac{%
2\pi k\sqrt{\left\langle n\right\rangle }}\lambda $ and whose format is
essentially that of $P_n.$ Therefore, for narrow enough $P_n$, some of the
terms $\omega _k\left( t\right) $ become completely disjoint from the rest,
describing an independent revival. Fleischhauer and Schleich were able to
use this formulation to derive many interesting results, such as the
decrease in amplitude and increase in width of each successive revival, and
also the number of revivals that can be resolved for a given state, before
they become scrambled due to their increasing width and consequent
interference with each other. Finally, they were also able to extend the
technique to some cases where $P_n$ is not smooth, such as in squeezed
states.

Using our dressed-state formulation for the population inversion, eq. (\ref
{popinv}), we are able to extend the Poisson Summation Formula method to
even more general initial states, including ones with atomic or atom-field
coherence. In the Appendix we show that, under suitable conditions, the
behaviour of the atomic inversion during its $k^{th}$ revival is described
by: 
\begin{equation}
\left\langle \sigma _z\left( t\right) \right\rangle \simeq \left( \frac{%
\lambda t}{\pi \sqrt{2k^3}}\right) \left. \left[ D\left( n\right) \cos
\left( \phi _n\left( 0\right) +\frac{\lambda ^2t^2}{2\pi k}-\frac \pi
4\right) \right] \right| _{n+1=\frac{\lambda ^2t^2}{4\pi ^2k^2}}
\label{kthrevival}
\end{equation}
where $D\left( n\right) $ is a continuous interpolation of the `weighted
dressedness' distribution $D_n$ of the initial state. Thus, in general it is
this distribution, not $P_n$, that is reflected in the shape of each
revival. In particular, in the case of population trapping, the quenching of
the revivals mirrors that of $D_n$. This explains the similarity between the
profiles in Figs. 2 and 3, including the doublet structure of the revivals 
in the limit of maximum trapping.

In order to display the accurateness of this expression, we use it to
calculate the revivals in the case of the Zaheer-Zubairy states $\left| \Psi
_{ZZ}\left( \alpha ,\gamma ,\xi \right) \right\rangle $ introduced above.
From expressions (\ref{zzdc}) for the dressed coordinates, we see that in
this case the weighted dressedness distribution is given by 
\begin{equation}
D_n\left( \alpha ,\gamma ,\xi \right) =\sqrt{Q_1^2\left( n\right) +\
Q_2^2\left( n\right) -2Q_1\left( n\right) \cdot \ Q_2\left( n\right) \cdot
\cos 2\left( \nu _\alpha -\xi \right) }  \label{Ddist}
\end{equation}
where 
\begin{mathletters}
\begin{eqnarray}
Q_1\left( n\right) &=&\exp \left( -\left| \alpha \right| ^2\right) \frac{%
\left| \alpha \right| ^{2n}}{\left( n\right) !}\cos ^2\left( \gamma \right)
\\
\ Q_2\left( n\right) &=&\exp \left( -\left| \alpha \right| ^2\right) \frac{%
\left| \alpha \right| ^{2(n+1)}}{\left( n+1\right) !}\sin ^2\left( \gamma
\right)
\end{eqnarray}
For sufficiently large $\alpha $, these Poissonian distributions are well
approximated by Gaussians: 
\end{mathletters}
\begin{mathletters}
\begin{eqnarray}
Q_1\left( n\right) &\simeq &\frac{\cos ^2\left( \gamma \right) }{\sqrt{2\pi }%
\left| \alpha \right| }\exp \left( \frac{-\left( n-\left| \alpha \right|
^2\right) ^2}{2\left| \alpha \right| ^2}\right) \\
Q_2\left( n\right) &\simeq &\frac{\sin ^2\left( \gamma \right) }{\sqrt{2\pi }%
\left| \alpha \right| }\exp \left( \frac{-\left( n+1-\left| \alpha \right|
^2\right) ^2}{2\left| \alpha \right| ^2}\right) .
\end{eqnarray}
Extending these to continuous values of $n$ and substituting in (\ref{Ddist}%
) (\ref{kthrevival}) we obtain an analytical expression for the $k^{th}$
revival in the inversion. In Fig. 5 we plot the first two of
these $(k=1,2)$ in the cases of maximum and minimum population trapping,
alongside the corresponding exact evolutions obtained numerically. Despite a
little distortion, agreement is seen to be very good.

A similar calculation is also possible in the case of the entangled
`even-odd' states $\left| \Psi _{EO}\left( \alpha ,\gamma ,\xi \right)
\right\rangle $ presented in the previous section. In this case, formula (%
\ref{kthrevival}) given above for the inversion is not valid, due to the
strong oscillations in $D_{n}$ [$D_{n}=0$ for odd $n$]. However, it is
possible to adapt our calculations to this situation (see section \ref
{varisec} in the Appendix), obtaining the expression:

\end{mathletters}
\begin{equation}
\left\langle \sigma _z\left( t\right) \right\rangle \simeq \left( \frac{%
\lambda t}{\pi \sqrt{k^3}}\right) \left[ D\left( m\right) \cos \left( \phi
_{2m}(0)+\frac{\lambda ^2t^2}{2\pi k}+\pi k-\frac \pi 4\right) \right]
\left| _{m+1=\frac{\lambda ^2t^2}{\pi ^2k^2}}\right.
\end{equation}
for the inversion during the $k^{th}$ revival. Here $D\left( m\right) $ is a
continuous interpolation of the even-indexed terms of $D_n$, renumbered with
the new index $m$ $(D\left( m=3\right) =D_6$, for instance). Comparing with
expression (\ref{kthrevival}) above for the Zaheer-Zubairy states, we can
see that the main difference in the present case is that the frequency of
revivals is {\it doubled}: the $k^{th}$ revival occurs around $%
t_k^{EO}\simeq \frac{\pi k\sqrt{\left\langle m\right\rangle +1}}\lambda $ ,
compared to $t_k^{ZZ}=\frac{2\pi k\sqrt{\left\langle n\right\rangle +1}}%
\lambda $ in the previous case. This difference is due essentially to the
doubling of the separation between adjacent Rabi frequencies in the present
case \cite{Antonio}, and is illustrated by comparing Figs. 2 and 4.

\section{Conclusion}

Intuitive pictures of the interaction between a two-level atom and an
electric field commonly involve the expectation that the atomic level
populations must change as both systems exchange excitations over the course
of time \cite{Eberly85}. This is due to the absence of further atomic
levels, which precludes the existence of destructive interference between
different atomic transitions. However, in a fully-quantised interaction
model such as the Jaynes-Cummings model, it is indeed possible to have
states in which the atomic populations are completely or nearly completely
trapped. This can be ultimately traced to the fact that the eigenstates of
this model are entangled. We have shown that, by giving up the traditional
point of view based on the individuality of each subsystem and assuming
instead one based on these entangled dressed states, it is possible to
obtain a quantitative understanding of population trapping in this model.
This is achieved via the introduction of a set of joint atom-field state
variables, the `weighted dressedness' distribution $D_{n}$ (eq. 
\ref{weidist}). We have shown that, for general initial conditions, this
distribution governs the evolution of the atomic inversion, assuming a role
commonly attributed to the photon number distribution $P_{n}$ (to which it
reduces in particular cases).\ Using $D_{n}$, we have obtained an upper bound
to the amplitude of population oscillations that can be expected from a
given state at any instant of its evolution. We have also been able to
obtain an approximate analytical description of the behaviour of the atomic
inversion during partially suppressed revivals. We have found that in general
the shape of revival envelopes is a direct reflection of the form of (a
continuous interpolation of ) $D_{n}$. In the particular case of a field
initially in a coherent state, this explains the appearance of a doublet
structure in the revivals in the limit of greatest population trapping.

\section{Acknowledgements}
We would like to thank Dr. S.M. Dutra and Prof. P.L. Knight for helpful  
comments and discussions. This work was supported by Conselho Nacional de 
Desenvolvimento Cient\'{i}fico e Tecnol\'{o}gico (CNPq) and Funda\c{c}\~{a}o 
de Amparo \`{a} Pesquisa do Estado de S\~{a}o Paulo (FAPESP), Brazil.

\begin{appendix}

\section{Approximate expressions for revivals in the atomic population
inversion for generalised initial conditions\label{AppendixA}}

In this Appendix, we calculate approximate expressions for the revivals in
the atomic population inversion, which are valid for a wide variety of {\it %
pure }initial conditions of the atom-field system. These calculations
generalise the method presented by Fleischhauer and Schleich \cite
{Fleischhauer93}, who assumed an initially factorised state of the form $%
\left| g\right\rangle _A\otimes \left| \psi \right\rangle _F.$ The result
they obtained for the $k^{th}$ revival is: 
\begin{equation}
\left\langle \hat{\sigma}_z\left( t\right) \right\rangle \simeq -P\left( n=%
\frac{\lambda ^2t^2}{4\pi ^2k^2}\right) \frac{\lambda t}{\pi \sqrt{2k^3}}%
\cos \left( \frac{\lambda ^2t^2}{2\pi k}-\frac \pi 4\right)
\label{Fleisresult}
\end{equation}
where $P_n$ is the photon distribution of state $\left| \psi \right\rangle
_C.$ In the case of more general initial conditions, including ones with
atomic and atom-field coherence, we shall see that the form of the revival
envelope depends not on the state's photon-number distribution, but on the
`weighted dressedness' distribution $D_n$ given in equation (\ref{weidist}).

As was shown in section \ref{inversec}, the inversion, written in terms of
dressed-state coordinates, has the form: 
\begin{equation}
\left\langle \hat{\sigma}_{z}\left( t\right) \right\rangle
=-w_{-1}+\sum_{n=0}^{\infty }D_{n}\cos \left( \phi _{n}\left( t\right)
\right)  \label{inversion0}
\end{equation}
where 
\begin{equation}
\phi _{n}\left( t\right) =\phi _{n}\left( 0\right) -\Omega _{n}t
\end{equation}

This expression can be rewritten according to the Poisson Summation Formula
(see Courant and Hilbert \cite{Courant53}, p.. 76 ) 
\begin{equation}
\left\langle \hat{\sigma}_z\left( t\right) \right\rangle =\sum_{k=-\infty
}^\infty \omega _k\left( t\right) +\tau _0\left( t\right)  \label{inversion}
\end{equation}
where 
\begin{mathletters}
\begin{eqnarray}
\omega _k\left( t\right) &=&\int_0^\infty dnD\left( n\right) \cos \left(
\phi _n\left( 0\right) -\Omega _nt\right) e^{2i\pi kn}  \label{wn} \\
\tau _0\left( t\right) &=&\frac 12D\left( 0\right) \cos \left( \phi _0\left(
0\right) -2\lambda t\right) -w_{-1}
\end{eqnarray}
and where $D\left( n\right) $ is {\it any }`reasonable' function of a
continuous variable $n$ (continuous, differentiable, etc.), that
interpolates between the values of $D_n$ at the points where $n$ integer.

Noting that the sum in $k$ in $\left( \ref{inversion}\right)$ extends to $%
\pm \infty $, so that the expression is invariant under $k\leftrightarrow -k$%
, it is possible to substitute eq. $\left( \ref{wn}\right)$ by:

\end{mathletters}
\begin{eqnarray}
\omega _k\left( t\right) &=&\int_0^\infty dnD\left( n\right) \cos \left(
\phi _n\left( 0\right) -2S_k\left( n,t\right) \right) = \\
&=&\int_0^\infty dnD\left( n\right) \cos \left( \phi _n\left( 0\right)
\right) \cos \left( 2S_k\left( n,t\right) \right) +  \nonumber \\
&&+\int_0^\infty dn\sin \left( \phi _n\left( 0\right) \right) \sin \left(
2S_k\left( n,t\right) \right) \\
&=&{Re}\int_0^\infty dnD_1\left( n\right) \exp \left( 2iS_k\left(
n,t\right) \right) +{Im}\int_0^\infty dnD_2\left( n\right) \exp \left(
2iS_k\left( n,t\right) \right)
\end{eqnarray}
where we have defined 
\begin{mathletters}
\begin{eqnarray}
S_k\left( n,t\right) &=&\pi kn-\lambda t\sqrt{n+1}  \label{Phase} \\
D_1\left( n\right) &=&D\left( n\right) \cos \left( \phi _n\left( 0\right)
\right)  \label{E1} \\
D_2\left( n\right) &=&D\left( n\right) \sin \left( \phi _n\left( 0\right)
\right) .  \label{E2}
\end{eqnarray}
In this way, the inversion may be rewritten (without any approximation) as

\end{mathletters}
\begin{equation}
\left\langle \hat{\sigma}_{z}\left( t\right) \right\rangle =\sum_{k=-\infty
}^{\infty }\omega _{k}^{1}\left( t\right) +\omega _{k}^{2}\left( t\right)
+\tau _{0}\left( t\right)  \label{inversionexp}
\end{equation}
where 
\begin{mathletters}
\begin{eqnarray}
\omega _{k}^{1}\left( t\right) &\equiv &{Re} \int_{0}^{\infty
}dnD_{1}\left( n\right) \exp \left( 2iS_{k}\left( n,t\right) \right) \\
\omega _{k}^{2}\left( t\right) &\equiv &{Im} \int_{0}^{\infty
}dnD_{2}\left( n\right) \exp \left( 2iS_{k}\left( n,t\right) \right)
\end{eqnarray}

Now, assuming the envelopes $D_1\left( n\right) $ and $D_2\left( n\right) $
are sufficiently smooth if compared to the oscillating functions $\cos
\left( 2S_k\left( n,t\right) \right) ,\sin \left( 2S_k\left( n,t\right)
\right) ,$ we may apply the method of stationary phases, approximating these
expressions by: 
\end{mathletters}
\begin{mathletters}
\begin{eqnarray}
\omega _k^1\left( t\right) &\simeq &D_1\left( n=n_k\right) {Re}\left\{ 
\begin{array}{c}
\exp \left( 2iS_k\left( n=n_k\right) \right) \times \\ 
\times \int_0^\infty dn\exp \left[ i\left. \frac{\partial ^2S_k}{\partial
n^2}\right| _{n=n_k}\left( n-n_k\right) ^2\right]
\end{array}
\right\}  \label{statphaprox} \\
\omega _k^2\left( t\right) &\simeq &D_2\left( n=n_k\right) {Im}\left\{ 
\begin{array}{c}
\exp \left( 2iS_k\left( n=n_k\right) \right) \times \\ 
\times \int_0^\infty dn\exp \left[ i\left. \frac{\partial ^2S_k}{\partial n^2%
}\right| _{n=n_k}\left( n-n_k\right) ^2\right]
\end{array}
\right\}
\end{eqnarray}

where $n_{k}$ is the point at which $\frac{\partial S_{k}}{\partial n}=0:$%
\end{mathletters}
\begin{equation}
n_{k}+1=\frac{\lambda ^{2}t^{2}}{4\pi ^{2}k^{2}}  \label{statpoint}
\end{equation}
We note that these expressions are invalid for $k=0,$ since in this case the
phase is always stationary (=0). Substituting in $\left( \ref{Phase}\right)
, $we obtain: 
\begin{eqnarray}
S_{k}\left( n=n_{k}\right) &=&-\left( \pi k+\frac{\lambda ^{2}t^{2}}{4\pi k}%
\right)  \label{S} \\
\left. \frac{\partial ^{2}S_{k}}{\partial n^{2}}\right| _{n=n_{k}} &=&2\frac{%
\pi ^{3}k^{3}}{\lambda ^{2}t^{2}}\equiv F  \label{K}
\end{eqnarray}

Now, the integral in $\left( \ref{statphaprox},b\right) $may be written in
terms of the Fresnel integrals \cite{Gradsteyn65}: 
\begin{mathletters}
\begin{eqnarray}
C\left( x\right) &=&\sqrt{\frac{2}{\pi }}\int_{0}^{x}dy\cos \left(
y^{2}\right) dy \\
S\left( x\right) &=&\sqrt{\frac{2}{\pi }}\int_{0}^{x}dy\sin \left(
y^{2}\right) dy
\end{eqnarray}
For instance, for $F>0:$%
\end{mathletters}
\begin{equation}
\int_{0}^{\infty }dn\exp \left( iF\left( n-n_{k}\right) ^{2}\right) =\sqrt{%
\frac{\pi }{2F}}\left[ C\left( x\rightarrow \infty \right) +C\left( \sqrt{F}%
n_{k}\right) +i\left( S\left( x\rightarrow \infty \right) +S\left( \sqrt{F}%
n_{k}\right) \right) \right]
\end{equation}
The asymptotic form of the Fresnel integrals for $x\rightarrow \infty $ \cite
{Gradsteyn65} is: 
\begin{mathletters}
\begin{eqnarray}
C\left( x\right) &\simeq &\frac{1}{2}+\sqrt{\frac{1}{2\pi }}\frac{\sin
\left( x^{2}\right) }{x}+O\left( \frac{1}{x^{2}}\right) \\
S\left( x\right) &\simeq &\frac{1}{2}+\sqrt{\frac{1}{2\pi }}\frac{\cos
\left( x^{2}\right) }{x}+O\left( \frac{1}{x^{2}}\right)
\end{eqnarray}
Assuming $\sqrt{\left| F\right| }n_{k}\gg 1$ and taking these approximations
to zeroth order, we have: 
\end{mathletters}
\begin{equation}
\int_{0}^{\infty }dn\exp \left( iF\left( n-n_{k}\right) ^{2}\right) \simeq 
\sqrt{\frac{\pi }{2F}}\left( 1+i\right) \text{ }(\text{valid for }F>0)
\label{asympt1}
\end{equation}
Similarly, for $F<0$%
\begin{equation}
\int_{0}^{\infty }dn\exp \left( iF\left( n-n_{k}\right) ^{2}\right) \simeq 
\sqrt{\frac{\pi }{2\left| F\right| }}\left( 1-i\right)  \label{asympt2}
\end{equation}
Using $\left( \ref{K}\right) $ and $\left( \ref{statpoint}\right) ,$%
condition $\sqrt{\left| F\right| }n_{k}\gg 1$ becomes: 
\begin{mathletters}
\begin{eqnarray}
\lambda t &\gg &2\left( \sqrt{\pi \left| k\right| }+\sqrt{2\pi \left|
k\right| +4\pi ^{2}k^{2}}\right)  \label{tcond} \\
\text{ou} &\ll &2\left( \sqrt{\pi \left| k\right| }-\sqrt{2\pi \left|
k\right| +4\pi ^{2}k^{2}}\right)
\end{eqnarray}
for $k=1,$ for example, this requires $\lambda t\gg 17$

Substituting the asymptotic expressions $\left( \ref{asympt1}\right) ,\left( 
\ref{asympt2}\right) $in $\left( \ref{statphaprox},b\right) ,$ we obtain for 
$k>0$ (or $k<0$):%
\end{mathletters}
\begin{mathletters}
\begin{eqnarray}
\omega _{k}^{1}\left( t\right) &\simeq &D_{1}\left( n=n_{k}\right) \sqrt{%
\frac{\pi }{2F}}\left[ \cos \left( \left. 2S_{k}\right| _{n=n_{k}}\right)
\mp \sin \left( \left. 2S_{k}\right| _{n=n_{k}}\right) \right] \\
\omega _{k}^{2}\left( t\right) &\simeq &D_{2}\left( n=n_{k}\right) \sqrt{%
\frac{\pi }{2F}}\left[ \cos \left( \left. 2S_{k}\right| _{n=n_{k}}\right)
\pm \sin \left( \left. 2S_{k}\right| _{n=n_{k}}\right) \right]
\end{eqnarray}
Thus, using $\left( \ref{K}\right) $ and $\left( \ref{inversionexp}\right) ,$%
the inversion may be written: 
\end{mathletters}
\begin{equation}
\left\langle \hat{\sigma}_{z}\left( t\right) \right\rangle =
\sum_{{k=-\infty, }{k\neq 0} }
^{\infty }\left( \frac{\lambda t}{2\pi k^{\frac{3}{2}}}%
\right) \left[ 
\begin{array}{c}
D_{1}\left( n=n_{k}\right) \left[ \cos \left( \left. 2S_{k}\right|
_{n=n_{k}}\right) \mp \sin \left( \left. 2S_{k}\right| _{n=n_{k}}\right)
\right] + \\ 
D_{2}\left( n=n_{k}\right) \left[ \cos \left( \left. 2S_{k}\right|
_{n=n_{k}}\right) \pm \sin \left( \left. 2S_{k}\right| _{n=n_{k}}\right)
\right]
\end{array}
\right] +\omega _{0}^{1}+\omega _{0}^{2}+\tau _{0}
\end{equation}
(where the upper(lower) sign is valid for the terms with $k>0(<0)$. Finally,
substituting the value $\left( \ref{S}\right) $ for $\left. S_{k}\right|
_{n=n_{k}}:$ 
\begin{eqnarray}
\left\langle \hat{\sigma}_{z}\left( t\right) \right\rangle &=&\sum_{{%
k=-\infty, }{k\neq 0} }^{\infty }\left( \frac{\lambda t}{2\pi k^{\frac{3}{2}}}%
\right) \left[ 
\begin{array}{c}
\left( \left. D_{1}+D_{2}\right| _{n=n_{k}}\right) \cos \left( 2\pi k+\frac{%
\lambda ^{2}t^{2}}{2\pi k}\right) \pm \\ 
\pm \left( \left. D_{1}-D_{2}\right| _{n=n_{k}}\right) \sin \left( 2\pi k+%
\frac{\lambda ^{2}t^{2}}{2\pi k}\right)
\end{array}
\right] +\omega _{0}^{1}+\omega _{0}^{2}+\tau _{0}  \nonumber \\
&=&\sum_{{k=-\infty, }{k\neq 0} }^{\infty }\left( \frac{\lambda t}{%
2\pi k^{\frac{3}{2}}}\right) \left. D\left( n\right) \right|
_{n=n_{k}}\left[ 
\begin{array}{c}
\left( \left. \cos \left( \phi _{n}\left( 0\right) \right) +\sin \left( \phi
_{n}\left( 0\right) \right) \right| _{n=n_{k}}\right) \cos \left( \frac{%
\lambda ^{2}t^{2}}{2\pi k}\right) \pm \\ 
\pm \left( \left. \cos \left( \phi _{n}\left( 0\right) \right) -\sin \left(
\phi _{n}\left( 0\right) \right) \right| _{n=n_{k}}\right) \sin \left( \frac{%
\lambda ^{2}t^{2}}{2\pi k}\right)
\end{array}
\right] +  \nonumber \\
&&+\omega _{0}^{1}+\omega _{0}^{2}+\tau _{0}
\end{eqnarray}

Finally, since: 
\begin{equation}
\left( \cos \left( x\right) +\sin \left( x\right) \right) \cos \left(
y\right) \pm \left( \cos \left( x\right) -\sin \left( x\right) \right) \cos
\left( y\right) =\sqrt{2}\cos \left( x\pm y-\frac \pi 4\right)
\end{equation}
then the approximate expression for the inversion is: 
\begin{equation}
\left\langle \hat{\sigma}_z\left( t\right) \right\rangle =\sum_{{%
k=-\infty, }{k\neq 0}}^\infty \left( \frac{\lambda t}{\sqrt{2}\pi k^{\frac 32}%
}\right) \left. \left[ D\left( n\right) \cos \left( \phi _n\left( 0\right)
\pm \frac{\lambda ^2t^2}{2\pi k}-\frac \pi 4\right) \right] \right| _{n+1=%
\frac{\lambda ^2t^2}{4\pi ^2k^2}}+\omega _0^1+\omega _0^2+\tau _0
\label{approxinversion}
\end{equation}
where the upper (lower) sign is valid for the terms with $k>0$ $(<0),$and
where $\tau _0$ and $\omega _0^1+\omega _0^2$ are given by 
\begin{mathletters}
\begin{eqnarray}
\tau _0\left( t\right) &=&\frac 12D\left( 0\right) \cos \left( \phi _0\left(
0\right) -2\lambda t\right) -w_{-1}  \label{aux} \\
\omega _0^1+\omega _0^2\left( t\right) &=&\omega _0\left( t\right)
=\int_0^\infty dnD\left( n\right) \cos \left( \phi _n\left( 0\right)
-2\lambda t\sqrt{n+1}\right)
\end{eqnarray}

The first of these two last terms represents the contribution to the
inversion of the states in which the field is in a vacuum state. For the
initial conditions which satisfy the approximations that have been made
above, this term will normally be negligible (see below). The second term
assumes non-negligible values only in the vicinity of $t=0$. This is due to
the fact that it is an integral over oscillating functions with different
frequencies, which rapidly cancel each other out; also, since these
frequencies form a continuum, they cannot re-phase substantially at
subsequent times. Fleischhauer and Schleich have thus conjectured that
zero-order terms of the Poisson Formula such as this always describe the
first {\it collapse }of the inversion shortly after $t=0$.

The remaining terms ( $k\neq 0$ ) each describe a modulated oscillation{\it %
\ }in the inversion, with an envelope given essentially by the shape of $%
D\left( n\right) $ and assuming non-negligible values only in an interval
around $t\simeq \frac{2\pi k\sqrt{\left\langle n\right\rangle +1}}\lambda $
(here $\left\langle n\right\rangle $ represents a value of $n$ around the
peak of $D\left( n\right) $). If $D\left( n\right) $ is sufficiently narrow,
the first few of these modulated oscillations will be well-separated in
time, thus constituting an independent {\it revival} during which the
inversion is described by 
\end{mathletters}
\begin{equation}
\left\langle \hat{\sigma}_z\left( t\right) \right\rangle \simeq \left( \frac{%
\lambda t}{\pi \sqrt{2k^3}}\right) \left. \left[ D\left( n\right) \cos
\left( \phi _n\left( 0\right) \pm \frac{\lambda ^2t^2}{2\pi k}-\frac \pi
4\right) \right] \right| _{n+1=\frac{\lambda ^2t^2}{4\pi ^2k^2}}
\end{equation}
When the initial state is of the form$\left| g\right\rangle _A\otimes \left|
\psi \right\rangle _C$ , we have 
\begin{equation}
\phi _n\left( 0\right) \rightarrow \pi \text{ },\text{ }D\left( n\right)
\rightarrow P_{n+1}
\end{equation}
( $P_n$ being the photon distribution of $\left| \psi \right\rangle _C)$, so
that Fleischhauer and Schleich's result (\ref{Fleisresult}) is recovered
(eq. $\left( 2.8b\right) $ in \cite{Fleischhauer93} ).

This expression is valid as long as:

\begin{enumerate}
\item[a) ]  The distributions $D_{n}\cos \left( \phi _{n}\left( 0\right)
\right) $ and $D_{n}\sin \left( \phi _{n}\left( 0\right) \right) $ vary
slowly with $n$ if compared with $\cos S_{k}\left( n,t\right) =\cos (\pi
kn-\lambda t\sqrt{n+1})$.

\item[b) ]  The value of $t$ obeys conditions $\left( \ref{tcond},b\right) .$
This implies that, for the $k^{th}$ term of the sum above to describe well
the $k^{th}$ revival, $D\left( n\right) $ must assume its largest values in
the region of $n$ where: 
\begin{equation}
n+1\gg \frac{4\left( 3\pi k+4\pi ^{2}k^{2}+2\pi k\sqrt{2+4\pi k}\right) }{%
4\pi ^{2}k^{2}}=4+\frac{1}{2\pi k}\left( \frac{3}{2}+\sqrt{2+4\pi k}\right) 
\end{equation}
Thus, the approximation can be expected to be good for all revivals if the
initial state has at least 10 or so photons on average in the field. This
will usually also ensure that the component $\tau _{0}$, which depends on $%
D\left( 0\right) $, (eq. (\ref{aux})), can be ignored.
\end{enumerate}

It is straightforward to show that in the examples of section IV,
where the dressed-state coordinates are given by expressions (\ref{zzdc})
and where $\left| \alpha \right| =7$, both of these conditions are satisfied.

\subsection{Variation for `even-odd' states [where $w_{2n-1}=0$]\label%
{varisec}}

In the case of `even-odd'-type states such as $\left| \Psi _{EO}\left(
\alpha ,\gamma ,\xi \right) \right\rangle $, where only the dressed-state
coordinates $w_{n}$ with {\it even }index are non-null $(w_{2n-1}=0),$
condition (a) above is violated and expression (\ref{approxinversion}) is
thus invalid. Nevertheless, a similar analytical expression for the
inversion can still be derived if eq. $\left( \ref{inversion0}\right) $ is
rewritten considering only the terms with even $n.$ In this case, it is
straightforward to show that, as long as the `continuous versions' of
distributions $D_{2n}\cos \left( \phi _{2n}\left( 0\right) \right) $ and $%
D_{2n}\sin \left( \phi _{2n}\left( 0\right) \right) $ are sufficiently
smooth {\it as a function of }${\it n}$, then the same stationary-phase
method as was used above can be applied, resulting in: 
\begin{equation}
\left\langle \hat{\sigma}_{z}\left( t\right) \right\rangle \simeq \sum_{%
{k=-\infty, }{k\neq 0}}^{\infty }\left( \frac{\lambda t}{\pi k^{\frac{3%
}{2}}}\right) \left. \left[ D\left( m\right) \cos \left( \phi _{2m}\left(
0\right) \pm \frac{\lambda ^{2}t^{2}}{2\pi k}+\pi k-\frac{\pi }{4}\right)
\right] \right| _{m+1=\frac{\lambda ^{2}t^{2}}{\pi ^{2}k^{2}}}\text{ }%
+\omega _{0}^{1}+\omega _{0}^{2}+\tau _{0}
\end{equation}
Here $D\left( m\right) $ is a continuous interpolation of the even-indexed
terms of $D_{n}$, renumbered with the new index $m$ $(D\left( m=3\right)
=D_{6}$, for instance). $\tau _{0}$ and $\omega _{0}^{1}+\omega _{0}^{2}$
are still given by $\left( \ref{aux},b\right) ,$except one must substitute $%
n\rightarrow m.$

Once again, it is possible to show that the approximations realised in the
course of obtaining this formula remain valid as long as $D\left(
m\right) $ assumes significant values only for $2m\gtrsim 10.$ (Thus, whenever
 the formula is applicable the term $\tau _0$ can be ignored).

\end{appendix}

\centerline{{\bf FIGURE CAPTIONS}} 
\vspace{0.2cm}

{\bf FIG. 1.} `Dressed-state coordinates' for
the Jaynes-Cummings model. Any state of the atom-field system can be
represented in terms of the set of parameters $\theta _{n},\phi
_{n},w_{n},\xi _{n}$ as defined in eqs.(\protect\ref{inista}) and (\protect\ref{dresscomp}).
For each $n$, the first two of these can be pictured as forming a Bloch-type
unit sphere, while the second two parametrise a circle of radius $w_{n}$.
Under the effect of the Jaynes-Cummings Hamiltonian, $w_{n}$ and $\theta _{n}
$ are constants of the motion, while $\phi _{n}$ and $\xi _{n}$ undergo
periodic motion at a frequency determined by the Rabi frequency $\Omega _{n}$.

{\bf FIG. 2.}Evolution of the atomic population inversion for an
initial state where the field is in a coherent state and the atom is in an
equally-weighed coherent superposition of $\left| e\right\rangle $ and 
$\left| g\right\rangle $, with relative phase $\xi $ (see eq. (\protect\ref{zzstate})). 
The amplitude of the coherent state is $\alpha =\left| \alpha \right|
e^{i\nu _{\alpha }}$, where $\left| \alpha \right| =7$ (49 photons on
average in the field). In (a), the relative phase $\left( \nu _{\alpha }-\xi
\right) $is equal to $\frac{\pi }{2}$, and the evolution follows the
familiar collapse-revival pattern. As this phase difference is lowered to
 $\frac{\pi }{10}$ (b) and finally zero (c), the revivals are quenched, and
the atomic populations become effectively trapped (notice the scale change
in (c) ). Simultaneously to this flattening, the revivals also assume a
doublet structure.

{\bf FIG. 3.} `Weighted dressedness' distributions $D_{n}=w_{n}^{2}\sin \theta _{n}$ 
for states where the field is in a coherent state $\left| \alpha \right\rangle $ with 
amplitude $\alpha=7e^{i\nu _{\alpha }}$ and the atom is in a coherent superposition 
$\cos\gamma \left| e\right\rangle +e^{-i\xi }\sin \gamma \left| g\right\rangle$.
In (a), $\gamma $ is fixed at $\frac{\pi }{4}$ (equal weights), while the
relative phase $(\nu _{\alpha }-\xi )$ varies from $\frac{\pi }{2}$ (I) to $%
\frac{\pi }{4}$ (II) to zero (III). The resulting flattening of $D_{n}$ is
due to the increasing dressedness of the components with the greatest
weights $w_{n}$. In (b), $(\nu _{\alpha }-\xi )$ is fixed at zero, while $%
\gamma $ varies from $\frac{\pi }{2}$ (1) to $\frac{\pi }{3}$ (2) to $\frac{%
\pi }{4}$ (III). In this case, the flattening corresponds to the matching of
the points $n_{\max }$ of maximum weight and $n_{\min }$ of maximum
dressedness (see eqs. \protect\ref{zzdc} ). The limit of maximum flattening 
corresponds to the greatest degree of population trapping in the time evolution. 
The appearance of a doublet structure in this limit is due to the most important 
component of the state becoming completely dressed $(\sin \theta _{n_{\max
}}\rightarrow 0)$. This is reflected in the shape of the quenched revivals,
as can be seen in Fig. 2.

{\bf FIG. 4.} Evolution of the atomic population inversion when the atom-field
system is initially in the entangled `even-odd' state $\left| \Psi
_{EO}\left( \alpha ,\gamma ,\xi \right) \right\rangle $ given in expression 
(\protect\ref{evenodd}). The parameters in (a)-(c) are the same as those in 
Fig. 2.
As in that case, a quenching of revivals and appearance of a doublet
structure is observed as $(\nu _{\alpha }-\xi )\rightarrow 0$. The main
difference in the present case is the halving of the interval between
adjacent revivals, due to the doubled spacing between the Rabi frequencies
present.

{\bf FIG. 5.} First and second revivals in the inversion for an initial state
where the atom is in a coherent superposition of $\left| e\right\rangle $
and $\left| g\right\rangle $ and the field is in a coherent state $\left|
\alpha \right\rangle $ with amplitude $\alpha =7$. The top two graphs are
close-ups of Figs. 2(a) and 2(c) and depict the numerical calculation of the
Jaynes-Cummings sum (\protect\ref{popinv}). The bottom two plot the approximate
analytical expression (\protect\ref{kthrevival}), obtained from the Poisson
Summation Formula. Despite a little distortion, agreement is seen to be
good. Note that, at the tail end of (a1) and (a2), the beginning of the
third revival can already be seen interfering with the second one, while in
(b1) and (b2) only the analytical expressions for first two revivals are
plotted.

%
  \begin{center}
    \leavevmode
    \epsfxsize=150mm
    \epsffile{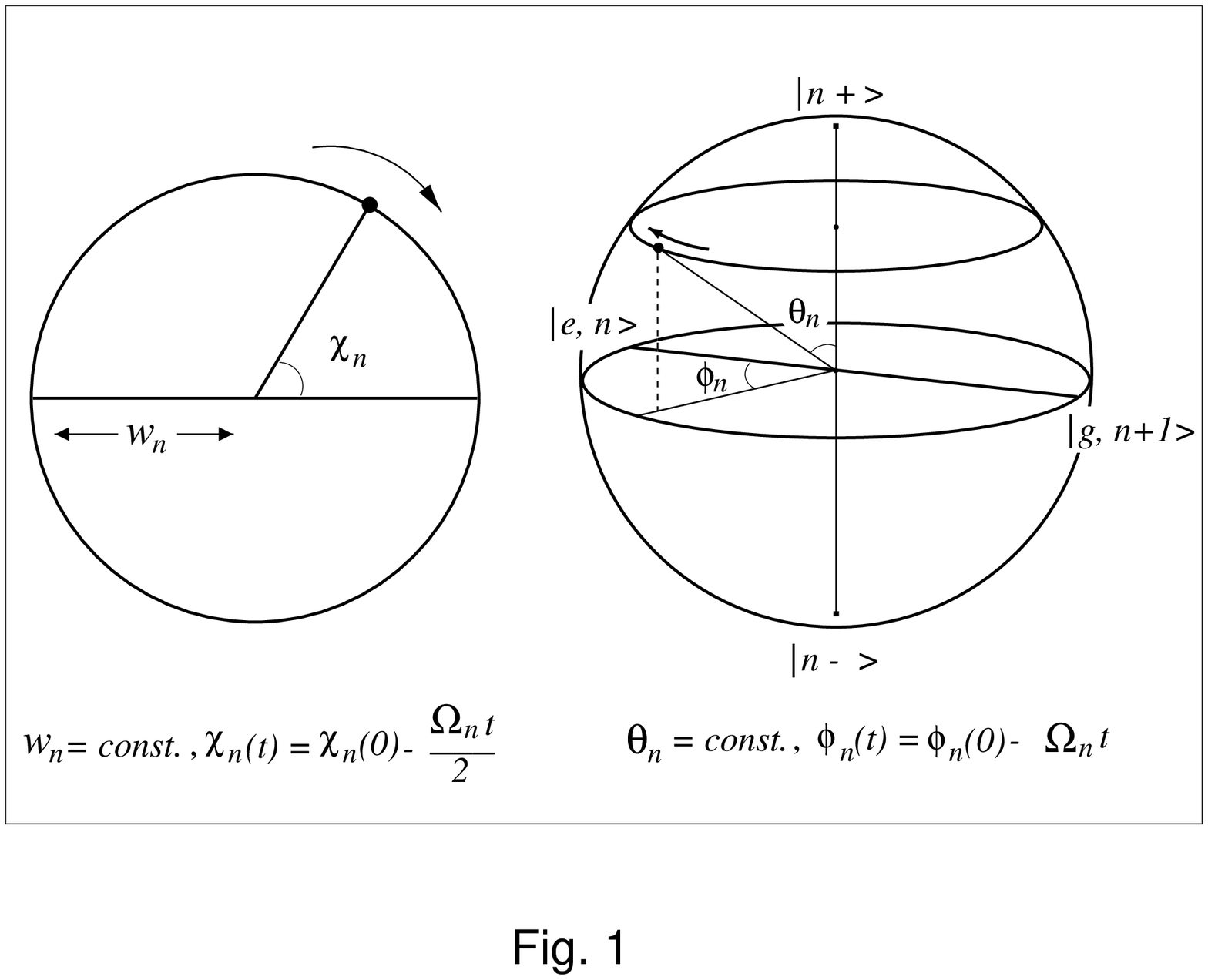}
  \end{center}

\newpage

  \begin{center}
    \leavevmode
    \epsfxsize=150mm
    \epsffile{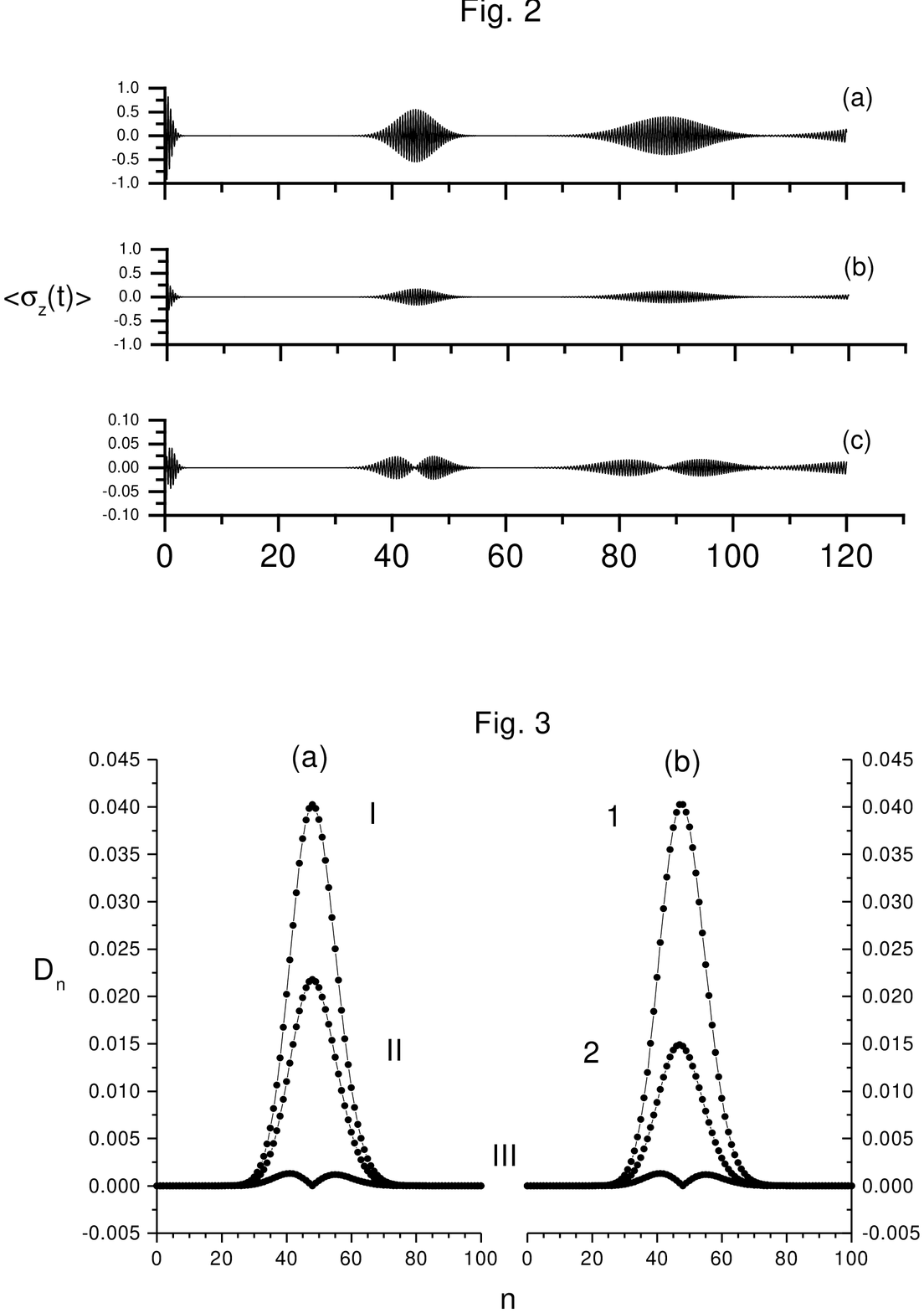}
  \end{center}

\newpage

  \begin{center}
    \leavevmode
    \epsfxsize=150mm
    \epsffile{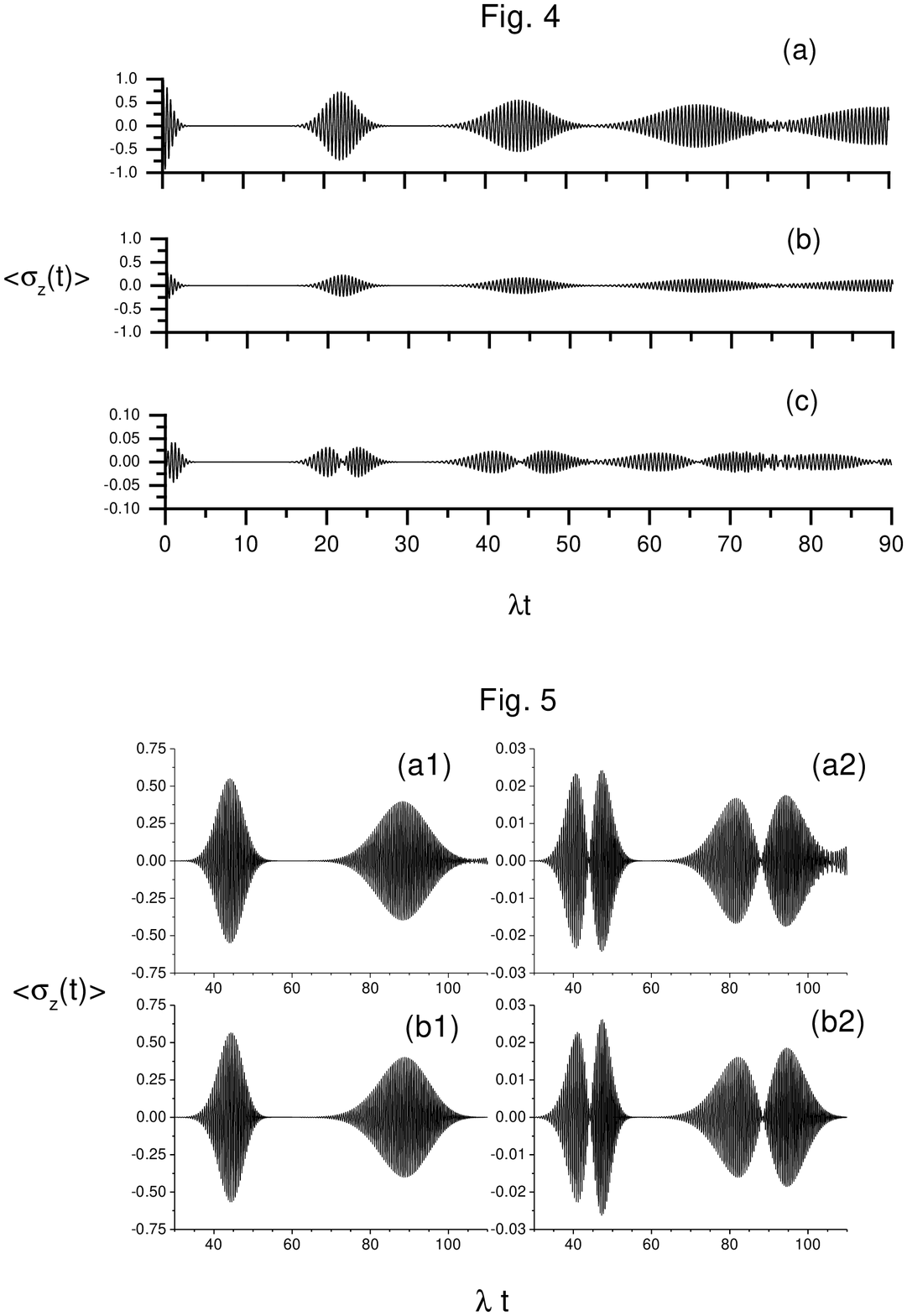}
  \end{center}


\begin{thebibliography}{99}
\bibitem{Jaynes63}  E.T. Jaynes and F.W. Cummings, Proc. IEEE {\bf 51}%
, 89 (1963).

\bibitem{Shore93}  B.W. Shore and P.L. Knight, J. Mod. Opt.{\bf 40},
1195 (1993).

\bibitem{Eberly80}  J.H. Eberly, N.B. Narozhny, and J.J.Sanchez-Mondragon, 
 Phys. Rev. Lett. {\bf 44}, 1323 (1980); N.B. Narozhny, J.J.
Sanchez-Mondragon, and J.H. Eberly, Phys. Rev. A {\bf 23}, 236 (1981);
H.I. Yoo, J.J. Sanchez-Mondragon and J.H. Eberly, J. Phys A {\bf 14 }%
,1383 (1981).

\bibitem{Phoenix88}  S.J.D. Phoenix and P.L. Knight, Ann. Phys. (N.Y.), 
{\bf 186}, 381 (1988);  Phys. Rev. A {\bf 44}, 6023 (1991); A. Ekert
and P.L. Knight, Am. J. Phys. {\bf 63}, 415 (1995).

\bibitem{Haroche94}  See for instance: S. Haroche and J.M. Raimond, in {\it %
Adv. At. Mol. Opt. Phys., Suppl. 2 }(Academic Press, New York, 1994) and references
therein.

\bibitem{Blockley92}  C.A. Blockley, D.F. Walls and H. Risken, 
Europhys. Lett. {\bf 17}, 509 (1992); D.M. Meekhof, C.Monroe, B.E. King,
W.M. Itano and D.J. Wineland, Phys. Rev. Lett. {\bf 76 },1796; {\bf 77}%
, 2346 (1996)

\bibitem{Brune96}  M. Brune, F. Schmidt-Kaler, A. Maali, J. Dreyer, J.M.
Raimond, and S. Haroche, Phys. Rev. Lett. {\bf 76}, 1800 (1996).

\bibitem{Fleischhauer93}  M. Fleischhauer and W.P. Schleich,  Phys. Rev.
A {\bf 47}, 4258 (1993).

\bibitem{Zaheer89}  K. Zaheer and M.S. Zubairy,  Phys. Rev. A {\bf 39},
2000 (1989).

\bibitem{GeaBanacloche91}  Julio Gea-Banacloche, Phys. Rev. A {\bf 44}%
, 5913 (1991).

\bibitem{Cirac90}  J.I. Cirac and L.L. S\'{a}nchez-Soto, Phys. Rev. A 
{\bf 42}, 2851 (1990).

\bibitem{Shapiro91}  J.H. Shapiro and S.R. Shepard, Phys. Rev. A {\bf %
43}, 3795 (1991).

\bibitem{Susskind64}  L. Susskind and J. Glogower, Physics, {\bf 1},
49 (1964).

\bibitem{Cirac91}  J.I. Cirac and L.L. S\'{a}nchez-Soto, Phys. Rev. A 
{\bf 44}, 3317 (1991).

\bibitem{Meystre91}  P. Meystre and M. Sargent III, Elements of Quantum
Optics, 2nd ed. (Springer-Verlag, Berlin, 1991).

\bibitem{Jonathan97}  D. Jonathan, M.Sc. thesis, Universidade
Estadual de Campinas, 1997.

\bibitem{Jonathan98}  D. Jonathan, K. Furuya and A.Vidiella-Barranco, in
preparation.

\bibitem{Hillery87}  A similar bound on the degree of deviation of the
field photon statistics from an initial Poissonian distribution has been
given in M.Hillery, Phys. Rev. A {\bf 35}, 4186 (1987).

\bibitem{remark}  We remark that expression (16a) given in \cite{Cirac90}
for the atomic state in this case is actually mistaken, and should be
replaced with the one given in expression (\ref{exact}) above.

\bibitem{remark2}  This phenomenon, which we address quantitatively in the
next section, is essentially due to the increased spacing between adjacent
Rabi functions. See \cite{Antonio}.

\bibitem{Antonio}  A. Vidiella-Barranco, H. Moya-Cessa and V. Buzek, J.
Mod. Opt. {\bf 39}, 1441 (1992).

\bibitem{Courant53}  R. Courant and D. Hilbert, {\it Methods of Mathematical
Physics, vol I} (Interscience , 1953).

\bibitem{Gradsteyn65}  I.S. Gradsteyn and L.M. Ryzhik, {\it Tables of
Integrals, Series and Products }(Academic Press, New York, 1965).

\bibitem{Eberly85}  H.-I. Yoo and J.H. Eberly, Phys. Rep. {\bf 118},
239 (1985).\newpage 
\end{thebibliography}
\end{document}